\documentclass[aps,prd,twocolumn,showpacs,amsmath,amssymb,nofootinbib,eqsecnum, preprintnumbers]{revtex4-2}
\usepackage{graphicx}
\usepackage{epstopdf}
\usepackage{epsfig}
\usepackage{xcolor}
\usepackage{setspace}
\usepackage{caption}
\usepackage[normalem]{ulem}  
\usepackage{soul} 
\setstcolor{red}  
\usepackage[tight,footnotesize]{subfigure}

\def\appendix{{\newpage\section*{Appendix}}\let\appendix\section%
        {\setcounter{section}{0}
        \gdef\thesection{\Alph{section}}}\section}

\newcommand{\be}{\begin{equation}}
\newcommand{\ee}{\end{equation}}
\newcommand{\bear}{\begin{eqnarray}}
\newcommand{\eear}{\end{eqnarray}}
\newcommand{\ba}{\begin{array}}
\newcommand{\ea}{\end{array}}

\begin{document}
\title{Configuration entropy and instability of accelerating black hole in AdS}
\author{Seoktae Koh}
\email{kundol.koh@jejunu.ac.kr}
\affiliation{Department of Science Education, Jeju National University, Jeju, 63243, Republic of Korea}
\author{Chong Oh Lee}
\email{cohlee@gmail.com}
\affiliation{Center of Liberal Arts, Wonkwang University, Iksan, Jeonbuk 54538, Republic of Korea}

\begin{abstract}
We consider an accelerating black hole with a negative cosmological constant in four-dimensional spacetime.
There are two configurations such as a black string (BS) phase when a mass parameter is zero
and a black hole (BH) phase when a mass parameter is non-zero.
We investigate their stability via the configuration entropy (CE).
It is found that the BS is not always stable but the BH has a thermally stable range
below the critical mass of the BH, which is consistent with thermodynamic instability.
\end{abstract}

\maketitle
\newpage

\section{Introduction}
The $p$-brane solutions have been suggested in~\cite{Gregory:1993vy, Gregory:1994bj} by adding a number of $p$-extra dimensions to
solutions for Einstein vacuum equations in higher dimensional spacetime and the simplest case of them is BS solutions by adding just one extra dimension in
such a spacetime. It has been shown that BS and $p$-brane in asymptotically flat higher dimensional spacetime are unstable
under gravitational perturbations~\cite{Gregory:1993vy, Gregory:1994bj} (see~\cite{Harmark:2007md} for reviews of the subject).
In particular, it has been shown that the BTZ BS with the negative cosmological constant in four-dimensional spacetime is unstable
in a certain range of the Kaluza-Klein (KK) mass under fermionic perturbation and gravitational perturbation~\cite{Liu:2008ds}
and a warped BS with nontrivial topologies in five-dimensional AdS spacetime is also unstable in some range of KK mass
under the linear gravitational perturbation~\cite{Yin:2010ix}.

Accelerating BH is described by the so-called $C$-metric~\cite{Kinnersley:1970zw, Plebanski:1976gy, Griffiths:2005qp},
which can be written in the solution to Einstein's equation with negative cosmological constant describing accelerating BH \cite{Plebanski:1976gy}.
It has been found that the special case of the $C$-metric~\cite{Emparan:1999wa} is a solution describing the BH
on the Planck brane in the Randall-Sundrum (RS) brane world model I~\cite{Randall:1999ee},
whose generalized version for the case of the AdS $C$-metric has been studied in~\cite{Emparan:1999fd}.
They have shown that there are two configurations such as the BS phase and the BH phase according to having or not having the mass parameter
and the BH has a thermally stable range below the critical mass of the BH.

The CE has suggested that the informational content in physical systems with localized energy density configurations
through measure of their ordering in field configuration space~\cite{Gleiser:2011di,Gleiser:2012tu}.
It has been extensively investigated for AdS/QCD holographic models~\cite{Bernardini:2016hvx,Bernardini:2016qit,Barbosa-Cendejas:2018mng,Karapetyan:2018yhm,Braga:2018fyc,Bernardini:2018uuy,daRocha:2021imz, daRocha:2021ntm, Karapetyan:2021ufz,Lee:2021rag,daRocha:2022bnk}
and performed for a detailed analysis of the thermal stability of a variety of physical systems~\cite{Gleiser:2012tu,Gleiser:2013mga,Gleiser:2014ipa,daRocha:2021imz,daRocha:2021ntm,
Karapetyan:2021ufz,Lee:2021rag,daRocha:2022bnk,Gleiser:2015rwa,Bernardini:2016hvx,Bernardini:2016qit,Casadio:2016aum,Braga:2016wzx,Lee:2017ero,Gleiser:2018kbq,Lee:2018zmp, Lee:2019tod, Barreto:2022ohl, Barreto:2022len, Barreto:2022mbx}.
Thus, it is intriguing that the issue of the stability is generalized to the accelerating BH with the negative cosmological constant in four-dimensional spacetime.

On the other hand, the Gubser-Mitra conjecture~\cite{Gubser:2000mm,Gubser:2000ec} has proposed
that the dynamical instability coincides with thermodynamic instability. In the last years attempt to test this conjecture for a BTZ BS
in four-dimensional spacetime was carried out in~\cite{Liu:2008ds} and was generalized to higher dimensional spacetime~\cite{Yin:2010ix}.
It is possible to investigate the stability of the BTZ BS solution via the CE. Therefore, it is also intriguing that such a conjecture is
established in this case.

The paper is organized as follows: In  section \ref{sect:accelBH}, we briefly review the accelerating BH with the negative cosmological constant in four-dimensional spacetime.
In section \ref{sect:ce}, we introduce the CE and explore thermodynamic instability for the BS and the BH. Finally, we summarize our results and  give our discussion in section \ref{sect:summary}.

\section{thermodynamics of accelerating BH}\label{sect:accelBH}
The metric with accelerating BH in $AdS_4$ \cite{Plebanski:1976gy} is given as
\bear\label{4aBHm}
ds_4^2=\frac{1}{A^2(x-y)^2}\left[P(y)dt^2-\frac{dy^2}{P(y)}+\frac{dx^2}{Q(x)}+Q(x)d\varphi^2\right]\nonumber\\
\eear
with
\bear
P(y)&=&-\lambda+ky^2-2mAy^3,\\
Q(x)&=&1+kx^2-2mAx^3,
\eear
and the discrete parameter $k$=+1, 0, $-1$. The metric (\ref{4aBHm}) satisfies the Einstein equation
\bear
R_{\mu\nu}=-\frac{3}{l_4^2}g_{\mu\nu},
\eear
where $l_4 =1/(A\sqrt{\lambda+1})$ is a four-dimensional (bulk) cosmological parameter with an acceleration parameter $A$  \footnote{Even if the parameter $A$ in the metric (\ref{4aBHm})
could be absorbed in a redefinition of
the coordinates, the four-dimensional cosmological parameter $l_4 =1/(A\sqrt{\lambda+1})$ sets
the scale for the four-dimensional constant $\Lambda_4=-3/l_4^2$ for $\lambda>-1$ and one can recover called the $AdS$ $C$-metric which describes black holes accelerating in $AdS_4$.
The parameter $A$ is related to acceleration of the black hole and it has been introduced for convenience.}
and parameter $\lambda$.
Here, the parameter $m$ has the relation with the mass of the BH and the constant $x/y$ slices have the branes~\cite{Emparan:1999wa, Emparan:1999fd}.
The metric (\ref{4aBHm}) has possible two configurations such as the BS and the BH.
(i) when $m=0$, adapting appropriate coordinate transformation, the BTZ BS in $AdS_4$ is written as~\cite{Emparan:1999fd}
\bear\label{4dBSm}
ds_4^2=l_4^2dz^2+a^2(z)l_4^2\left[-(\lambda r^2-k)dt^2+\frac{dr^2}{\lambda r^2-k}+r^2d\varphi^2\right]\nonumber\\
\eear
with $a(z)=\sqrt{\lambda}\cosh(z)$. Since the metric (\ref{4dBSm}) on the constant $z$ slices under $\lambda>0$ has locally the geometry of $AdS_3$,
it for $k=+1$ is reduced to the geometry of a BTZ BH~\cite{Banados:1992wn} on surfaces of constant $z$
\bear\label{3dBSm}
ds_3^2=-(\lambda r^2-1) dt^2+\frac{dr^2}{\lambda r^2-1}+r^2 d\varphi^2,
\eear
which satisfies $R_{AB}=-2\lambda g_{AB}$. Here, the capital Latin indices $A$ and $B$ refer to the brane coordinates 0, 1, 2, and the Greek indices $\mu$ and $\nu$
to the bulk coordinates 0, $\ldots$, 3.
(ii) when $m>0$, the metric function $Q(x)$ has exactly one positive root $x=\alpha$, and is set with $\Delta\varphi=4\pi/|Q'(x=\alpha)|$ to avoid a conical singularity at $x=\alpha$.
The BH localized on a brane is given as~\cite{Emparan:1999fd}
\bear\label{3dBHm}
ds_3^2&=&\frac{1}{A^2}\left[-\left(\lambda r^2-k-\frac{2mA}{r}\right)dt^2\right.\nonumber\\
&&+\left.\frac{dr^2}{\lambda r^2-k-\frac{2mA}{r}}+r^2d\varphi^2\right],
\eear
which is similar to the above AdS metric~(\ref{3dBSm}), but has the extra term $2mA/r$ which is caused by the four-dimensional nature of the BH.
In particular, in the case of $k=+1$ when $2mA$ is very much smaller than $1/\sqrt{\lambda}$, the extra term is negligible outside the horizon, and
the exterior geometry is exactly matched with that of the BTZ BH. However, when $2mA$ is very much bigger than $1/\sqrt{\lambda}$,
there will be significant deviations from the BTZ metric due to outside the BH~\cite{Emparan:1999fd}.
Furthermore, it has been shown that after introducing RS brane world model I~\cite{Randall:1999ee},
four-dimensional newton's constant $G_4$ has the relation with three-dimensional newton's constant $G_3$ as follows:
\bear\label{G3}
G_3=\frac{AG_4}{2}.
\eear
Employing a single auxiliary variable $z$, the parameter $2mA$ may be defined as
\bear
2mA=\frac{(\lambda+z^2)z\sqrt{1+z}}{(\lambda-z^3)^{3/2}}.
\eear

As mentioned before, the metric (\ref{4dBSm}) on the constant $z$ under $\lambda>0$ for $k=+1$ reduces to the metric (\ref{3dBSm}).
Since we will focus on the metric (\ref{3dBSm}), it is convenient to set $\lambda=2$ from now on.

\begin{figure}[!htbp]
\begin{center}
{\includegraphics[width=7cm]{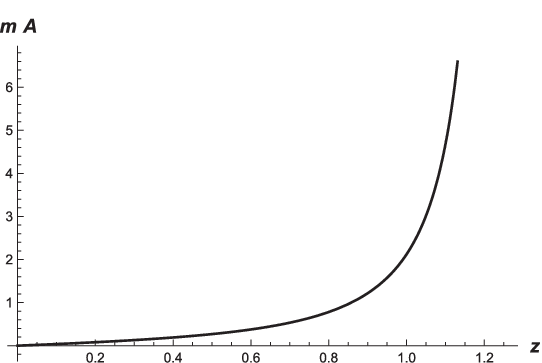}}
\end{center}
\vspace{-0.6cm}
\caption{{\footnotesize From Eq. (2.9), plot of the parameter $mA$ as the function of $z$ for $\lambda=2$.}}
\label{figI}
\end{figure}

As you see in Fig. 1, the parameter $mA$ increases monotonically as the parameter $z$ grows up.
When $mA$ goes to zero, $z\sim 2mA\sqrt{\lambda}\sim 0$ while when $mA$ becomes infinite, $z\sim\lambda^{1/3}$ ($z\sim\sqrt[3]{2}=1.25992$ for $\lambda=2$).

Using the first law of BH thermodynamics, the three-dimensional mass of the BTZ BH on the brane $M_3$ can be expressed in terms of the variable $z$
\bear\label{M3}
M_3=\frac{1}{2G_3}\frac{z^2(1+z)(\lambda-z^3)}{(\lambda+3z^2+2z^3)^2},
\eear
and the four-dimensional mass of the BTZ BS $M_4$ yields also
\bear\label{M4}
M_4=\frac{1}{G_4 A}\frac{z^2(1+z)(\lambda-z^3)}{(\lambda+3z^2+2z^3)^2},
\eear
which is exactly matched with the three-dimensional mass $M_3$ (\ref{M3}) by taking Eq. (\ref{G3})~\cite{Emparan:1999wa, Emparan:1999fd}.

\begin{figure}[!htbp]
\begin{center}
{\includegraphics[width=7cm]{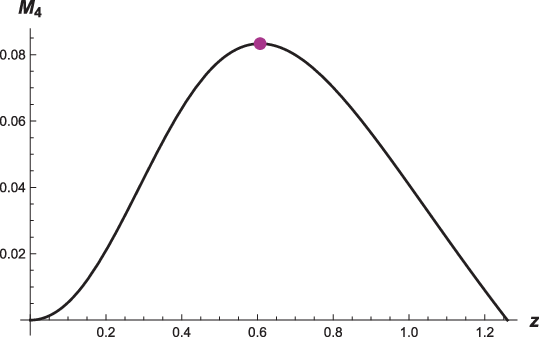}}
\end{center}
\vspace{-0.6cm}
\caption{{\footnotesize From Eq. (2.11), plot of the four-dimensional mass of the BTZ BS $M_4$ as the function of $z$ for $A=1$, $\lambda=2$ and $G_4=1$.}}
\label{figII}
\end{figure}

As shown in Fig. 2, for $A=1$, $\lambda=2$, and $G_4=1$ the maximum mass (the purple point in Fig. 2) at
$z=\frac{1}{4}(\sqrt[3]{15+4\sqrt{14}}+\sqrt[3]{15-4\sqrt{14}}-1)$
is given as
\bear\label{Mmax4}
M_{\rm 4,max}=\frac{1}{12}.
\eear
Then, it has been found that the four-dimensional entropy of the BS is able to be given as
the three-dimensional entropy of the BTZ BH on the brane
\bear\label{S3}
S_{\rm BS}=\frac{\pi}{A}\sqrt{\frac{2M_3}{\lambda G_3}},
\eear
and the four-dimensional entropy of the BH in the bulk is
\bear\label{S4}
S_{\rm 4, BH}=\frac{2\pi}{G_4 A^2}\frac{z}{\lambda+3z^2+2z^3}.
\eear
For $A=1$, $\lambda=2$, and $G_4=1$ the maximum entropy (the purple point in Fig. 3) at
$z=\frac{1}{4}(\sqrt[3]{15+4\sqrt{14}}+\sqrt[3]{15-4\sqrt{14}}-1)$
is given as
{\tiny
\bear\label{Smax4}
S_{\rm 4,max}&=&
\frac{16\pi\left(\sqrt[3]{15+4\sqrt{14}}+\sqrt[3]{15-4\sqrt{14}}-1\right)}{64+\left(\sqrt[3]{15+4\sqrt{14}}+\sqrt[3]{15-4\sqrt{14}}-1\right)^2
   \left(\sqrt[3]{15+4 \sqrt{14}}+\sqrt[3]{15-4 \sqrt{14}}+5\right)}\nonumber\\
   &=&1.07354.
\eear
}
As mentioned earlier, $z$ approaches to $\lambda$ as the parameter $mA$ goes to infinity ($M_4\rightarrow0$).
The four-dimensional entropy of the BH $S_{\rm BS}$ ({\ref{S4}) has the finite limit at $z=\lambda^{1/3}$
\bear
S_0=\frac{2\pi}{3G_4 A^2}\frac{1}{\lambda^{1/3}(\lambda^{1/3}+1)},
\eear
which for $A=1$, $\lambda=2$, and $G_4=1$ becomes
\bear\label{S0}
S_0=\frac{2 \sqrt[3]{2} \pi }{\sqrt[3]{4} \left(3+2 \sqrt[3]{2}\right)+2}=0.735567,
\eear
(the orange point in Fig. 3) at $z=2^{1/3}$.
Thus, the BH entropy $S_{\rm 4, BH}$ (\ref{S4}) (the blue dotted curve in Fig. 4) grows up from 0 at $mA=0$ ($z=0$ and $M_4=0$)
to the maximum value $S_{\rm 4,max}$ (\ref{Smax4}) at $mA=2/\sqrt{27}$ ($z=\frac{1}{4}(\sqrt[3]{15+4\sqrt{14}}+\sqrt[3]{15-4\sqrt{14}}-1)$ and $M_4=M_{\rm 4,max}=\frac{1}{12}$),
and then it (the green solid curve in Fig. 4) decreases to the finite limit $S_0$ (\ref{S0}) at $mA=\infty$ ($z=\lambda^{1/3}$ and $M=0$)

\begin{figure}[!htbp]
\begin{center}
{\includegraphics[width=7cm]{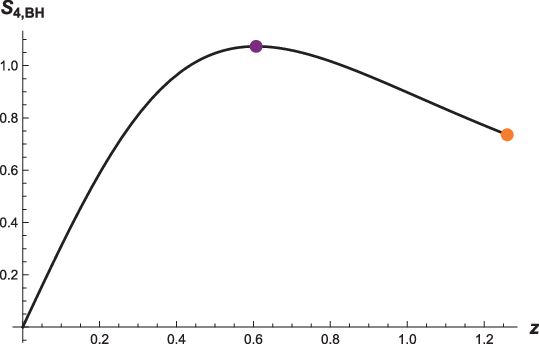}}
\end{center}
\vspace{-0.6cm}
\caption{{\footnotesize From Eq. (2.14), plot of the four-dimensional entropy  of the BH in the bulk $S_{\rm 4, BH}$
as the function of $z$ for $A=1$, $\lambda=2$ and $G_4=1$.}}
\label{figIII}
\end{figure}

Upon substituting  the three-dimensional mass of the BTZ BH $M_3$ (\ref{M3}) and the relation
of $G_3$ and $G_4$ (\ref{G3}) into the BTZ BS (\ref{S3}), one can get
\bear\label{z1}
z(z^3+z^2-\lambda)=0,
\eear
through the matching between the entropy of the BS (\ref{S3}) and that of the BH (\ref{S4}).
When $z=0$, the mass and entropy of the BH with small $mA$  (the blue dotted curve in Fig. 4)
are equal to zero, which matches that of the BS (the green solid curve in Fig4.) as you see in Fig 4.
One considers the matching between the mass and entropy of the BS (the green solid curve in Fig4.)
and that of the BH with large $mA$ (the red dotted curve in Fig4.), it satisfies
\bear\label{z2}
z^3+z^2=\lambda.
\eear
The root of Eq. (\ref{z2}) for $\lambda=2$ is $z=1$ which denote the crossover (black) point in Fig. 4.
Then, the critical entropy $S$ and the critical mass $M$ are given as
\bear\label{Sc}
S_c&=&\frac{2\pi}{7}=0.897598,\\
\label{Mc}
M_c&=&\frac{2}{49}=0.0408163.
\eear
As shown in Fig. 4, two branches (the red dotted curve and the blue dotted curve) are joined at the maximum mass $M_{\rm 4,max}$ (the purple point in Fig. 4).
It has been investigated to be the instability of two localized BH solutions and the BS by employing entropy~\cite{Emparan:1999fd}.
When $0\leq M\leq M_{\rm 4,max}$ ($0\leq mA \leq 2/\sqrt{27}$), the BS entropy (the green solid curve in Fig. 4) and the BH entropy (the blue dotted curve in Fig. 4)
increase as the mass grows up. Furthermore, the BS entropy (the green solid curve in Fig. 4)
is bigger than the BH entropy (the blue dotted curve in Fig. 4). It means that the BS is thermally stable in such region.
When $2/\sqrt{27}~(M= M_{\rm 4,max})\leq mA \leq \infty~(M=0)$, the BH entropy decreases to finite value $S_0$ (\ref{S0}) as the parameter $mA$ grows up. In particular,
the BH entropy (the red dotted curve in Fig. 4) is bigger than
the BS entropy (the green solid curve in Fig. 4) below the critical value $mA=3\sqrt{2})$ ($M=M_c$ (\ref{Mc})), and it becomes stable.
\begin{figure}[!htbp]
\begin{center}
{\includegraphics[width=7cm]{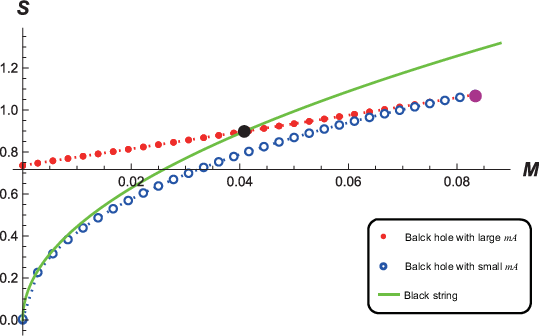}}
\end{center}
\vspace{-0.6cm}
\caption{{\footnotesize Upon substituting the three-dimensional mass of the BTZ BH $M_3$ (2.10) and the relation of $G_3$ and $G_4$ (2.8) from Eq. (2.13), plot of the entropy of the BS of
the mass and plot of the entropy of the BH of the mass by the results of the numerical calculation from Eq. (2.11) and Eq. (2.14) for $A=1$, $\lambda=2$ and $G_4=1$.}}
\label{figIV}
\end{figure}

\begin{center}
\begin{table*}[hbt!]
\begin{center}
\caption{The CE of BH and BS for the various masses.}
\end{center}
{\scriptsize
\begin{tabular}{ |c|r|r|r|r|r|r|r| }
\hline
$M$    &$S_{C,{\rm BH,L}}$ &$S_{C,{\rm BS,up}}$ &$S_{C,{\rm BH,S}}$      &$S_{C,{\rm BS,low}}$     &$\Delta S_c$     &$\Delta\bar{S}_c$      &$\Delta\tilde{S}_c$\\
\hline
$10^{-6}$   &879908.46641&1079714.23102&880121.49299&880121.49299&213.02657&199805.76461 &0          \\
0.00277874  &16692.18597 &17435.76435  &16696.18108 &14212.64119 &3.99510  &743.57838    &-2483.53989\\
0.00555649  &11804.22174 &12268.01277  &11807.01267 &10000.18469 &2.79092  &463.79103    &-1806.82798\\
0.03888942  &4461.96612  &4461.97115   &4462.80813  &3637.14453  &0.84200  &0.00502      &-825.66360 \\
2/49        &4355.37492  &4355.37492   &4356.18100  &3550.25336  &0.80603  &0            &-805.92759 \\
0.04444491  &4173.80821  &4155.74900   &4174.55060  &3387.53001  &0.74239  &-18.05881    &-787.02059 \\
0.07777784  &3213.13131  &2990.26234   &3155.47560  &2437.49140  &0.22063  &-164.99263   &-717.98420 \\
0.08055559  &3155.25497  &2845.61707   &3100.56033  &2319.58483  &0.15373  &-254.78952   &-780.97550 \\
1/12        &3048.36909  &2645.35510   &3048.36909  &2156.34268  &0        &-403.01399   &-892.02641 \\
\hline
\end{tabular}
}
\end{table*}
\end{center}

\begin{figure*}[hbt!]
\subfigure[]{\includegraphics[width=7cm]{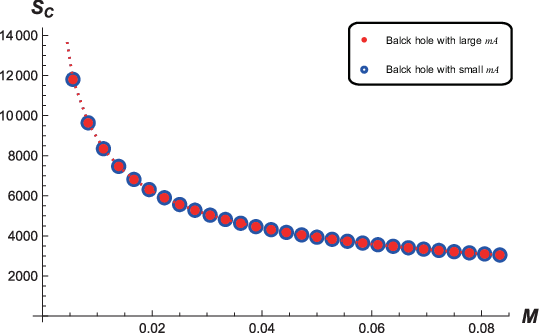}}~~~~~
\subfigure[]{\includegraphics[width=7cm]{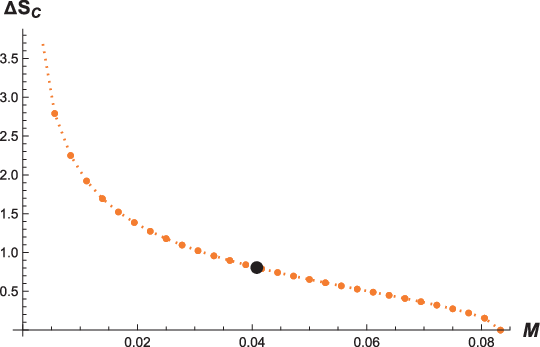}}
\caption{\footnotesize{(a) From the Table 1, plot of the BH CE as a function of the mass $M$ for $A=1$, $\lambda=2$ and $G_4=1$.
The blue circles denote the BH CE from $M=10^{-6}$ ($mA=0.001$) to $M=1/12$ ($mA=2/\sqrt{27}$)
and the red points does it from $M=10^{-6}$ ($mA=3.7037\times 10^7$) to $M=1/12$ ( $mA=2/\sqrt{27}$).
(b) From the Table 1, plot of  the CE difference of the BH with small $mA$ and the BH with large $mA$ as a function of the mass $M$
from $M=10^{-6}$ to $M=1/12$, which always has positive values except for $M=1/12$ ($\Delta S_C =0$). Here, the black point denotes the critical mass $M_c$.}}
\end{figure*}

\begin{figure*}[hbt!]
\subfigure[]{\includegraphics[width=7cm]{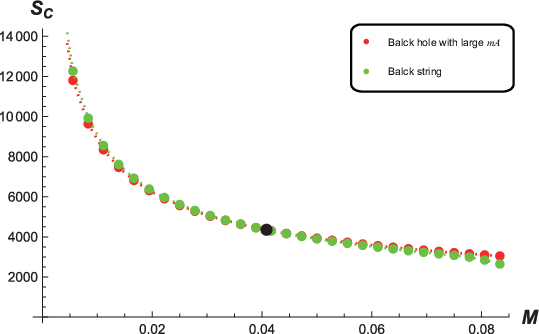}}~~~~~
\subfigure[]{\includegraphics[width=7cm]{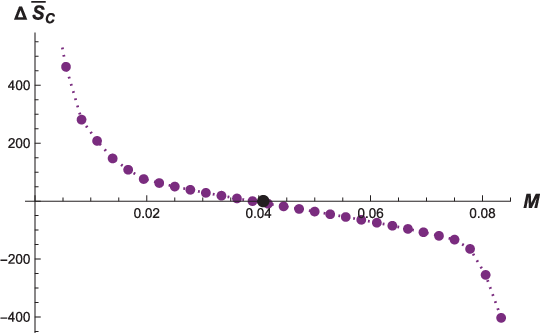}}
\caption{\footnotesize{(a) From the Table 1, plot of the BH CE and the BS CE as a function of the mass $M$ for $A=1$, $\lambda=2$ and $G_4=1$.
The red and green points respectively denote the BC CE and BS CE from $M=10^{-6}$ ($mA=3.7037\times 10^7$) to $M=1/12$ ($mA=2/\sqrt{27}$).
(b) From the Table 1, plot of  the CE difference of the BS and the BH with large $mA$ as a function of the mass $M$
(from $M=10^{-6}$ to $M=1/12$), which always has negative values beyond the critical mass $M_c=2/49$ ($\Delta\bar S_C =0$). Here, the black points denote the critical mass $M_c$.}
}
\end{figure*}

\begin{figure}[!htbp]
\begin{center}
{\includegraphics[width=7cm]{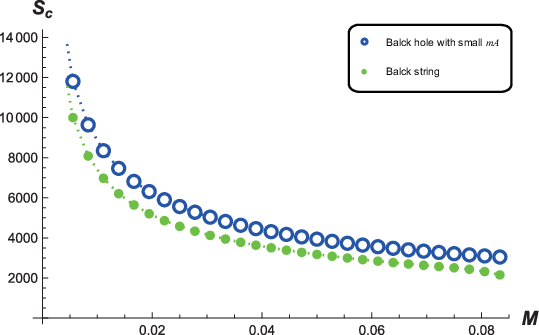}}
\end{center}
\vspace{-0.6cm}
\caption{{\footnotesize From the Table 1, plot of the BH CE and the BS CE as a function of the mass $M$ for $A=1$, $\lambda=2$ and $G_4=1$.
The blue and green points respectively denote the BH CE and BS CE from $M=10^{-6}$ ($mA=3.7037\times 10^7$) to $M=1/12$ ($mA=2/\sqrt{27}$).}}
\label{figIV}
\end{figure}
\section{CE and Instability for the BTZ BS and BH} \label{sect:ce}
As discussed in the previous section, the four-dimensional mass of the BTZ BS $M_4$ (\ref{M4}) in the bulk is exactly matched with the three-dimensional
mass $M_3$ (\ref{M3}) on the brane via Eq. (\ref{G3}). From the metric (\ref{3dBSm}), the energy density of the BS $\rho_{\rm BS}$ is given as
\bear\label{BSrho}
\rho_{\rm BS}\equiv T_{00}=A^3 \lambda  \left(4 {G_4} {M_3}-A \lambda  r^2\right),
\eear
and from the metric (\ref{3dBHm}), the energy density of the BH $\rho_{\rm BH}$ is
\bear\label{BHrho}
\rho_{\rm BH}\equiv T_{00}=\frac{2 m^2 A^2}{r^4}+\frac{{m A} \left(\lambda  r^2+1\right)}{r^3}+\lambda -\lambda ^2 r^2.\nonumber\\
\eear
There is the system with spatially localized energy in position space $x$ and the Fourier transform of the energy density $\rho(x)$
as the function of the position $x$ at frequency $k$ is given as the complex number

\bear
\rho(k)=\frac{1}{\sqrt{2\pi}}\bigg(\alpha(k)+i\beta(k)\bigg),
\eear
with
\bear
\alpha(k)&=&\int_{0}^{x_h}\rho(x)\cos(k x)dx,\\
\beta(k)&=&\int_{0}^{x_h}\rho(x)\sin(k x)dx,
\eear
where there is an event horizon at $x=x_h$.
Then, the modal fraction of $\rho(k)$ is defined as~\cite{Gleiser:2018kbq}
\bear
{\cal F}(k)=\frac{|\rho(k)|^2}{|\rho(k)|_{\rm max}^2},
\eear
with $|\rho(k)|^2=\alpha^2(k)+\beta^2(k)$
where $|\rho(k)|_{\rm max}^2$ denotes the maximum value of $|\rho(k)|^2$.
One may define the CE~\cite{Gleiser:2012tu} as
\bear\label{CE}
S[{\cal F}]=-\int_{-\infty}^{\infty}{\cal F}(k)\log[{\cal F}(k)]dk.
\eear

From now on, considering the CE difference of the BS and the BH, we will investigate the
instability of the BS/BH system.

The energy density of the BS (\ref{BSrho}) is independent of the parameter $mA$
while that of the BH (\ref{BHrho}) is depended on it.
However, in the case of the upper branch (the red dotted curve in Fig. 4), the entropy of the BS matches that of the BH at the crossover (black) point in Fig. 4,
and in the case of the lower branch (the blue dotted curve in Fig. 4), the entropy of the BS matches that of the BH at $M=0$ in Fig. 4.
Thus, the CE of the BS should match two branches of the BH at such points.

After adapting these matching conditions and the CE (\ref{CE}),
from the energy densities (\ref{BSrho}) and (\ref{BHrho}), the CE of the BS and that of the BH
are numerically obtained as shown in Table I. The CE of the BH with large/small $m A$ is written as $S_{C,{\rm BH,L}}$ and  $S_{C,{\rm BH,S}}$, and
that of the BS with matching condition at the upper/lower branch of the BH is $S_{C,{\rm BS,up}}$ and  $S_{C,{\rm BS,low}}$.
The CE dfference of $S_{C,{\rm BH,L}}$ and $S_{C,{\rm BH,S}}$, that of $S_{C,{\rm BH,L}}$ and $S_{C,{\rm BS,up}}$,
and that of $S_{C,{\rm BH,S}}$ and $S_{C,{\rm BS,low}}$ are respectively defined as
$\Delta S_C\equiv S_{C,{\rm BH,S}}-S_{C,{\rm BH,L}}$, $\Delta{\bar S}_C\equiv S_{C,{\rm BS,up}}-S_{C,{\rm BH,L}}$, and
$\Delta \tilde{S}_C\equiv S_{C,{\rm BS,low}}-S_{C,{\rm BH,S}}$ as shown in Table I and plot them as you see in Fig. 5, Fig. 6, Fig. 7.

As shown in Fig. 5. (a), the CE of the BH with large/small $mA$ monotonically decreases as the mass $M$ grows up.
In particular, since the CE difference of $S_{C,{\rm BH,L}}$ and $S_{C,{\rm BH,S}}$, $\Delta S_C$, is always positive as shown in Table I and in Fig. 5. (b),
the CE of the BH with large $m A$  $S_{C,{\rm BH,L}}$ is stable, which correctly coincides
with the result of thermodynamic analysis in section \ref{sect:accelBH} that the BH with large $mA$ is thermally stable
through the comparison of the entropy of the BH with large $mA$ and that of the BH with small $mA$ for given mass $M$
as you see the red dotted curve and the blue dotted curve in Fig. 4.

Considering the CE difference of $S_{C,{\rm BH,L}}$ and $S_{C,{\rm BS,up}}$, $\Delta{\bar S}$,
as the mass $M$ decreases (the parameter $mA$ increases), the BH has a smaller CE below the critical mass $M_c$ defined in
(\ref{Mc}) (at which $\Delta {\bar S}_C =0$) as you see in  Table I and Fig. 6.
The BH becomes thermally stable in this region, which is consistent with that in~\cite{Emparan:1999fd}

As shown in Table I and Fig. 7, for the BH with small $mA$,the CE difference of $S_{C,{\rm BS,low}}-S_{C,{\rm BH,S}}$, $\Delta{\tilde S}$ is always negative except
for $M=10^{-6}$ (at which $\Delta{\tilde S}=0$),
and the BS is thermally stable, which is well matched with that in~\cite{Emparan:1999fd}.
On the other hand, it has been explored dynamical instability against gravitational/tensor perturbation for the BS~\cite{Liu:2008ds,Yin:2010ix}.
It has been found that the BS is stable beyond the critical value against perturbation, while it is unstable below the critical value.
These instability behaviors which  qualitatively agrees with thermodynamic instability behaviors in~\cite{Emparan:1999fd}
also supports the Gubser-Mitra conjecture of that the dynamical instability coincides with thermodynamic instability.

\section{Summary and Discussion} \label{sect:summary}
The accelerating BH in four-dimensional AdS spacetime has two configurations such as the BS phase and the BH phase.
We explored the thermal instability of these two configurations via CE.
The CE of the BS and that of the BH were numerically obtained as shown in Table I.
As the mass $M$ increases, the CE of the BH with large/small $mA$ monotonically decreases as well as that of the BS decreases
as shown in Fig. 5. (a), Fig. 6. (a), and Fig. 7.
It was found that the BS is thermally unstable (the BH is stable) below the critical mass $M_c$ (\ref{Mc})
when the BH has large $mA$ while the BS is always
stable when the BH has small $mA$, which is well matched with that in~\cite{Emparan:1999fd}.

Recently it has been proposed that  for accelerating black holes in AdS spacetime they have calculated the dual stress-energy tensor for the spacetime
and identify the energy density associated with a static observer at infinity. It has been shown that the dual energy-momentum tensor is able to be given
by a non-hydrodynamic contribution with a universal coefficient to a three-dimensional perfect fluid and the result of the holographic computation is well matched
with that of conformal completion method for the mass~\cite{Anabalon:2018ydc}.
It is intriguing to investigate whether the thermal instability from both sides yields the same result through employing the CE.

$\newline$
$\newline$

{\it Data Availability Statements}: the data sets generated during and/or analyzed during the current study are available from the corresponding author upon reasonable request.

\section*{Acknowledgements}
This work was supported by Basic Science Research Program through the National
Research Foundation of Korea (NRF) funded by the Ministry of Education, Science and
Technology (NRF-2021R1A2C1005748 (SK), NRF-2018R1D1A1B07049451 (COL)).

\end{document}